**Coherent ultrafast spin-dynamics probed in three dimensional topological insulators**

[1,2]F. Boschini, [1]M. Mansurova, [3]G. Mussler, [3]J. Kampmeier, [3]D. Grützmacher, [4]L. Braun, [5]F. Katmis, [5]J. S. Moodera, [6]C. Dallera, [6]E. Carpene, [7]C. Franz, [7]M. Czerner, [7]C. Heiliger, [4]T. Kampfrath, [8]M. Münzenberg

1. I. Physikalisches Institut, Georg-August-Universität Göttingen, Friedrich-Hund-Platz 1, 37077 Göttingen, Germany

2. Dipartimento di Fisica, Politecnico di Milano, 20133 Milan, Italy

3. Peter Grünberg Institut (PGI-9) and Jülich-Aachen Research Alliance (JARA-FIT), Forschungszentrum Jülich, 52425 Jülich, Germany

4. Department of Physical Chemistry, Fritz Haber Institute, Faradayweg 4-6, 14195 Berlin, Germany

5. Massachusetts Institute of Technology, Cambridge, Massachusetts 02139, USA

6. IFN-CNR, Dipartimento di Fisica, Politecnico di Milano, 20133 Milan, Italy

7. I. Physikalisches Institut, Justus-Liebig-Universität Gießen, 35392 Gießen, Germany

8. Institut für Physik, Ernst-Moritz-Arndt Universität Greifswald, Felix-Hausdorff-Straße 6, 17489 Greifswald, Germany

**Topological insulators are candidates to open up a novel route in spin based electronics. Different to traditional ferromagnetic materials, where the carrier spin-polarization and magnetization are based on the exchange interaction, the spin properties in topological insulators are based on the coupling of spin- and orbit interaction connected to its momentum. Specific ways to control the spin-polarization with light have been demonstrated: the energy momentum landscape of the Dirac cone provides spin-momentum locking of the charge current and its spin. The directionality of spin and momentum, as well as control with light has been demonstrated. Here we demonstrate a coherent femtosecond control of spin-polarization for states in the valence band at around the Dirac cone.**

When exciting materials with femtosecond laser radiation, generally the observed phenomena are manifold. The photon energy is directly transferred to electrons. The sharp electronic energy distribution smears out and the initially coherent excited electronic system thermalizes by scattering processes in the momentum space resulting in decoherence. Since electrons carry a spin, magnetic information is strongly entangled with this coherence – decoherence transition process. The rearrangement of charge in addition causes a movement of the atoms, however with a delay of picoseconds. In ferromagnetic materials, a coherent interaction of the photon field with the spins has been evidenced that triggered research in the new direction of femtomagnetism [1,2,3]. As one mechanism the generation of a photo-induced optical anisotropy in oxides like $YMnO_3$ has been discussed. This effect is related to a distinct localized level scheme and acts as a mechanism to create an ultrashort spin polarization [4].

Non-equilibrium spin-polarization in semiconductors [5] leads to spin-galvanic effects [6] and/or spin helical modes due to an interplay of spin coherence and diffusion [7]. Here, very interesting is the case of topological insulators [8, 9, 10] that combines both: spin-properties, via the spin momentum locking [11, 12, 13, 14, 15, 16, 17, 18, 19, 20], and semiconductor properties because of their band gap. In the presence of the laser field further coherent effects emerge, shown in [21], which arise from an interaction of the periodic laser field and the electron wave functions of the surface bands leading to laser-field induced gaps with periodicity of $\hbar\omega$ in energy, known as Floquet Bloch states. The question investigated in this study is, what happens to a spin-polarization induced by the near infrared laser field and is it possible to generate a polarization induced by the photon field that interacts coherently with the spins in the topological insulator? In the following we disentangle the different processes by femtosecond pump-probe experiments giving access to their dynamics. The different stages manifest from the fingerprint of a magnetic coherent state during the first ten femtoseconds of the laser excitation to coherent atomic motion beyond the picosecond time scale originating from strain pulses, seen as GHz oscillation in the magneto-optical Kerr rotation. Possibly this will allow to manipulate carrier spin polarized carrier dynamics interacting with the surface bands in the gap of the topological insulator, and the spin dynamics in the spin-polarized surface states itself on THz frequencies.

To probe the ultrafast dynamics in the topological insulator at the surface within the optical penetration depth, simultaneously to the time-resolved reflectivity, the change in the probe-pulse polarization is determined [22]. This allows sampling not only the electron and phonon dynamics, but also transient birefringence induced by the pump-pulse, and eventual rearrangements of the spin polarization that can be induced in first tens of femtosecond to nanoseconds after laser excitation [23,24]. If a transient spin-imbalance is present in the all-optical pump-probe experiment the probe pulse's polarization related to a complex Kerr angle of the probe beam can be detected. In our experiments the pulse has a full width of 40-80 fs and a wave length of $\lambda = 800$ nm corresponding to 1.55 eV photon energy. The complex polarization angle (rotation of the polarization axis and ellipticity) is extracted using an optical-modulation technique (see methods). All experiments have been conducted at 295 K, the Kerr angles are in the order of a few 1/100 degree and comparable to ultrafast demagnetization experiments.

In Fig. 1 we present the phonon modes indicating the typical phonon resonance as a proof of the high quality of the $Bi_2Te_3$ thin film layered samples surface. The movement of the atoms is traced by the delayed femtosecond probe in the reflectivity evolution after pump-pulse excitation. This phonon mode at 1.84 THz represents a 540 fs dynamics of the atoms moving in the $A_{1g}$ mode, which is the most prominent mode [25,26]. On a longer time scale a slower oscillation is evident. It originates from the thermal expansion within the surface after energy deposition, triggering a breathing of the $Bi_2Te_3$ layer. This phonon frequency is determined by the thickness and the velocity of sound ~4nm/ps (4000 m/s) and we extract a breathing of the standing wave of 63 GHz for the longitudinal phonon mode.

An ultrafast access to the pump-pulse induced change of birefringence is given by the complex polarization rotation presented in Fig. 2. It is governed by the dielectric tensor ε, which (in the polar configuration) has a particularly easy form with a diagonal non-magnetic part and an off-diagonal part that is related to a complex polarization angle $\phi_K = -i\, \epsilon_{xy}/(\epsilon_{xx}-1)\sqrt{\epsilon_{xx}}$ and present on different time scales [27]. In ferromagnetic materials the magneto-optical Kerr angle is proportional to the materials magnetization. Moreover, it is known that the normalization of the spin-sensitive off-diagonal optical contribution with the non-magnetic diagonal optical contribution can cause a non-magnetic modification of the optical properties, changed by the modification of the optical reflectivity. A direct comparison of the transient Kerr signal with the time-resolved reflectivity gives a first hint whether the signal constitutes the pure magnetic signal or the diagonal elements of the dielectric tensor cross talk with the complex Kerr angle, as the crystal is rotated [28] or as a function of delay time [29]. In non magnetic materials different effects can contribute to a change in the complex polarization rotation which is sensitive to off-diagonal terms of the dielectric tensor. We present the pump-pulse induced changes in the complex polarization angle of the probe pulse as a function of the delay time in Fig. 2 b for different orientation of the crystal with respect to the plane of incidence of the laser light. Interestingly one observes a 30-ps slow oscillation period, corresponding to a 31 GHz mode, decaying with a rate of 100 ps. This oscillation's amplitude mirrors the crystal symmetry of the film while rotating the crystal: the film growth on the substrate is defined by the underlying symmetry. Two domains are known to be present, rotated by 180° to each other [20]. This converts the hexagonal symmetry of the crystal that nominally would give rise to a three-fold symmetry with period of 120°, to a two-fold symmetry with 180° period seen in Fig. 2 d. Furthermore it can be noted that the Kerr rotation is not sensitive to the phonon modes at 63 GHz, since those are not present in the Kerr signal. Experiments at different magnetic fields and reversing the handedness of the circular polarization of the pump beam does not provide evidence for spin-packets propagating at the sample surface, as in the case of the laser induced spin helix in a GaAs/AlGaAs quantum well for example [7]. Such a signal should reverse the spin orientation of the pump-pulse induced spin polarization. This 31 GHz mode is not present in the dynamics of the reflectivity where we find the longitudinal standing wave. A coupling of the off-diagonal optical elements of the optical tensor to a different breathing mode of alternative symmetry cannot be excluded as a probable origin here.

A dependence on the pump-pulse polarization is seen in the presence of the pump-pulse's laser field on ultrashort time scales. It can be separated as shown in Fig. 2 a, inset. The Kerr rotation for right circularly and left circularly polarized pump pulse reveals a characteristic negative or positive peak, while the linear pump-pulse excitation shows an intermediate slope. The latter contribution is related the heat and stress induced effects, as observed for the longer delay times, and suggests discussing the difference signal in the following. We notice that this ultrafast signal remembers the light-field polarization. It seems related to a pump-pulse induced ultrashort time polarization in the material, evident in a sign change of the signal. Due to its fast decay within excitation pulses's duration, it seems to be coherent within the excitation pulse at first glance. Such an ultrafast polarization induced signal in the refraction has been observed by Gedik et al. and ascribed to an ultrafast induced spin polarization followed by a very fast depolarization [14]. By studying Kerr rotation and ellipticity simultaneously, we will see in the following that this *ad hoc* explanation of the shape of the signal is not correct.

To determine its properties we conducted a composition dependent experimental series shifting the Fermi energy though the gap of the topological insulator $(Bi_{1-x}Sb_x)_2Te_3$ alloy film with varied Sb content (39, 43 and 45 %). The variation goes from n-doped via intrinsic to p-doped. For Sb=43% a minimum carrier density is extracted from Hall effect measurements of n ~ $10^{17}$ cm$^{-3}$ (in the following compensated sample). To extract a spin-signal related to a pump-pulse induced spin-asymmetries at ultrafast time scales we use the pump right circular, left circular pump-probe Kerr rotation and ellipticity as a difference signal to the linear pump-pulse polarization. The linear pump-pulse induced changes are assumed to be spin-polarization independent. It appears that both signals, presented in Fig. 3, have a characteristic slope: the shape of the signal is a combination of a Gaussian like or a Gaussian-derivative like signal. Moreover the ellipticity has a significant delay as compared to the Kerr rotation, which amounts to about 80 fs. The sign change for the Kerr rotation is observed as small hump at a positive delay, e.g. later arrival of the probe-beam pulse relative to the pump-beam's center. The opposite holds for the Kerr ellipticity where it appears at negative delay, e.g. earlier arrival of the probe-beam pulse relative to the pump-beam's center and points to a phase shift in-between the signals. To extract the time scales of the dynamics and its components we employ a model applied originally to GaAs [5] and adapted for photo-induced circular optical anisotropy in oxide insulators [4]. There it describes the dipole transition for excitation above the band gap, from hybrid 2p-3d ground states into $Mn^{3+}$ 3d states where within a characteristic time (Raman coherence time $\tau_R$) with two ingredients, a transition in a degenerate state with symmetry $|+j\rangle$ to $|-j\rangle$ changing magnetic quantum number and thus the magnetic polarization. This allows describing the typical shape of the polarization observed of the complex Kerr rotation ($\phi_K = \theta + i\varepsilon$) around a resonance

$$\theta + i\varepsilon = A \exp\left(-\frac{t^2}{\sigma^2}\right) + B \exp\left(\frac{\sigma^2}{4\tau_R^2} - \frac{t}{\sigma}\right) \times \left(1 - \text{erf}\left(\frac{\sigma}{2\tau_R} - \frac{t}{\sigma}\right)\right) \qquad (1)$$

with *t* as temporal evolution (pump-probe delay) and $\sigma = \tau_{FWHM}/\sqrt{2ln2}$, the pulse width is given by $\tau_{FWHM}$ full width at half maximum (FWHM), Raman coherence time $\tau_R$ and the complex amplitudes are related to the dielectric susceptibility tensor in analogy to Ref. [4].

|  | $(Bi_{1-x}Sb_x)_2Te_3$ | | |
| --- | --- | --- | --- |
|  | n-doped, Sb=39% | Intrinsic, Sb=43% | p-doped, Sb=45% |
| **Raman coherence time (fs)** | 9.0(0.5)-13.7(1.2) | 8.9(2)-9.8(0.1) | 11(3)-14.8(1.0) |

Table 1: Raman coherence times $\tau_R$ extracted from the time-resolved Kerr rotation and ellipticity spectra for different doping $(Bi_{1-x}Sb_x)_2Te_3$, Sb=39, 43, 45 %. The analysis was tested against rigidity. The interval reflects the spread using fixed and variable length of the excitation pulse respectively (statistical error in brackets).

It is noted that the pump-induced anisotropy is quite large compared to the amplitudes of the signal at longer time scales. The Kerr rotation and ellipticity induced by the pump-pulse for oblique incidence of 45° from the normal were fitted simultaneously. The signal is asymmetric, and as a consequence we extract a Raman coherence time that is shorter than the laser-pulse width. We find almost constant values around 9-15 fs practically independent of the Fermi level position, with a tendency for a minimum at Sb=43%. The amplitude factors, related to time-dependent dielectric tensor elements, are responsible for the phase factor. One can see the effect of the different amplitudes given in Fig. 4 a, where the ellipticity $i\varepsilon$ in the complex Kerr rotation follows almost a Gaussian shape, and we generate an almost derivative shape for the real part $\theta$. This demonstrates that a coherent polarization effect determines the shape of the spectra in presence of the femtosecond laser field. The phenomenological three level model can be used to derive the different shape observed for Kerr rotation and Kerr ellipticity, where both the real and imaginary part are related via the Kramers-Kronig relation. The analysis is shown in Fig. 3 overlaid to the experimental data and describes the spectra in detail. In Fig. 4 a the comparison between real part and imaginary part is displayed for the case of the compensated sample for one circular polarization of the pump pulse (right polarized) and linearly polarized probe pulse and in Fig. 4 d the complex Kerr signal is compared for right and left circularly excitation showing the opposite direction of the phase-trace in the complex plane.

To relate these transitions to the band structure of $Bi_2Te_3$ (or $Sb_2Te_3$) we use *ab initio* density functional theory band structure calculations and solve the full Dirac equation; the electron-electron interaction is treated in local-density approximation. The bulk-band structure of $Bi_2Te_3$ (or $Sb_2Te_3$) at around the Fermi level is dominated by bonding p-orbital of the Te and the anti-bonding of Bi in a simplified level scheme. We calculated the j resolved band structure in the simplified model for j=1/2 and j=3/2 contributions. For the laser frequency $\omega$ with $\hbar\omega = 1.55\ eV$ we find a resonance for dipole transitions for $\sigma^-$ from $m_j$=3/2 to $m_j$ =1/2 ($\sigma^+$ from $m_j$ =-3/2 to $m_j$ =-1/2) at the Γ point into the lowest states unoccupied states above the energy gap at ~ 100 meV and the ground state is at 1.5 eV. The final states are related to the dominant symmetry $|+1/2\rangle$ and $|-1/2\rangle$. Relatively flat bands at around the Γ point result in a reasonably large localized density of states for these transitions to form a resonance, corresponding to a transition from Te to the anti-bonding of Bi verified by band structure calculations (see Supplementary Materials). This polarization effect induced change in the Kerr rotation appears only in presence of the pump-pulse. After that the coherence is lost, originating from intra- and interband scattering and dephasing. Thus, this probe of the femtosecond phase space filling is a sensitive probe of the time dependent redistribution of carriers excited at the band edge close to the Γ point and is sensitive to the electronic structure at the Dirac cone here modified by the different composition of the $(Bi_{1-x}Sb_x)_2Te_3$ alloy.

By relating these transitions to the element projected band structure (see Supplementary Materials), we can interpret them as rectification and shift currents [30, 31]. While the first one results in a diffusive current with different velocities of the spin-polarized excited carriers, the other results in a replacement of charge (spin polarized) in the unit cell and a short polarization. In the simplified level scheme of $Bi_2Te_3$ this corresponds a transfer of electrons from the bonding p-orbital of Te to the anti-bonding p-orbital of Bi. Indications for this local shift currents that are hence polarization dependent are seen also in THz emission and pulsed AC photo-voltage generation in Hall bar devices and such current burst can be related to a spin polarized charge current [32]. In contrast, the Kerr angle allows accessing the pump-induced coherent spin-polarization in the transient state directly. It is interesting to relate our results to the spin- and electron dynamics determined by time-resolved angle-resolved spectroscopy at the same transition by Cacho et al. [33]. They report a second spin polarized surface state in the conduction band. From their data and experiments they conclude a well separated spin dynamics in these excited states with significant spin polarization is present, however the effects they observe appear on a much longer time scales of 1-6 ps. They identify this second spot to contribute with a major role to the spin dynamics of the excited system.

In conclusion we discussed dynamics effects induced by the pump pulse at 1.55 eV laser femtosecond laser excitations detected in the reflectivity dynamics and complex Kerr signal on time scales of femtosecond to nanoseconds. Coherent phonon modes have been observed, as the $A_{1g}$ mode and modes in the GHz range related to acoustic standing waves of the film. Modes in this frequency range have been observed in the Kerr rotation, e.g. the dynamic signal related to the magnetic properties, however we can exclude a direct, simple crosstalk of optics from the comparison of both modes that have different origin and interaction may be more complex. By subtraction of the linear-pump pulse dynamics from left- and right circular pump-pulse polarization, we find a spin-related signal present only during the laser excitation with different phase for real and imaginary part of the complex Kerr angle. This signal is only present due to polarization of the topological insulator in the presence of the pump-pulses' light field. We describe the shape of these signals with a pump-induced polarization of the states excited across the band gap of the topological insulator, involving the upper edge of the conduction band that can be analyzed in terms of a Raman coherence time. These states are hybridized with the surface bands and it will be interesting to relate these pump induced polarization effect to the properties of the topological states in the band gap in future studies. We find a weak dependence for the Raman times with the tendency of a shorter coherence time for the intrinsic films with intermediate Sb composition.

**Acknowledgements**

T.K, C.H., G.M.,J.K., D.G., and M.M. thank the Deutsche Forschungsgemeinschaft (DFG) for funding within project "Investigation of directional THz spin currents in topological surface states", Schwerpunktprogramm 1666 „Topological Insulators: Materials – Fundamental Properties – Devices". G.M., J.K., and D.G. also thank the Virtual Institute of Topological Insulator (VITI) for financial support.

**Methods**

*Ultrafast laser spectroscopy*

In the all-optical pump probe experiment the magneto-optical Kerr rotation of the probe beam is measured. The pulse has a full width at half maximum of $\tau$ = 40-80 fs and a central wave length of $\lambda$ = 800 nm (Ti:Sapphire amplifier system RegA 9040, Coherent) corresponding to 1.55 eV photon energy. The repetition rate of the laser system used is 250 kHz. The Kerr angle is extracted by means of a photoelastic modulator. All measurements have been conducted at room temperature (295 K). To extract a spin-signal related to a pump pulse induced spin-asymmetries at ultrafast time scales, we use the Kerr signal for linear pump-pulse polarization as a non-magnetic reference. Since the linear pump-pulse induces only transitions equal for spin-up and spin downs, these changes will be spin-independent. The signals for pump right circular, left circular, Kerr rotation and ellipticity are plotted as a difference signal pump-left-, right circular minus pump-pulse linearly polarized (see inset Fig 2 a). This allows separating the contributions from electron and phonon dynamics and pump-induced spin-polarization. Interference of scattered light of pump-pulse and probe-pulse at the detector caused by stray light from the pump region, for example by a locally rough spot or periodic modulation of the index of refraction (light induced optical grating), have been identified in the experiments and minimized. This effect known as optical artifact at around $\tau$=0 is causing a fast oscillation of the signal sampled at each delay. The analysis using the three level model was tested against rigidity of the interdependence of the two variables defining the interval in table 1 reflect the spread using fixed and variable length of the excitation pulse respectively.

*Sample preparation*

The first set of samples, $Bi_2Te_3$ and $(Bi,Sb)_2Te_3$ films from the Forschungszentrum Jülich were grown by molecular beam epitaxy (MBE) on Si(111) wafers. Prior to the deposition, the Si substrates were chemically cleaned by the HF-last RCA procedure to remove the native oxide and passivate the surface with hydrogen. The substrates were subsequently heated in-situ to 600°C for 20 min to desorb the hydrogen atoms from the surface. The Sb, Bi and Te material fluxes were generated by effusion cells with temperatures of 460°C-470°C (Bi), 370°C-390°C (Te), and 400°C-420°C (Sb). For all samples, the Te shutter was opened 2 seconds before the Sb and/or Bi shutter, in order to saturate the Si substrate surface with Te.

Throughout the growth, the substrate temperature was set at 300°C. A low growth rate of ~ 5 - 10 nm/h was chosen in order to obtain a smooth and uniform sample surface. In case of the ternary $(Bi,Sb)_2Te_3$ samples, the Sb concentrations were determined by Raman measurements and x-ray photoelectron spectroscopy. The growth of $Bi_2Se_3$ layers at the Massachusetts Institute of Technology, second set of films studied, was carried out by molecular beam epitaxy (MBE) under an ultra-high vacuum (UHV) environment ($10^{-9}$-$10^{-10}$ Torr). High purity (5N) elemental Bi and Se were co-evaporated from different Knudsen-cells with typical growth rates between 0.5-1 nm/min. (0001)-oriented, epi-ready, commercial $Al_2O_3$ (sapphire) wafers were used as a substrate. To improve the surface quality common wet cleaning techniques were employed, followed by *in-situ* oxygen plasma cleaning to remove organic contaminants from the substrate, finally annealing for an hour at 600 °C, and at 800 °C for 30 min under $10^{-9}$ Torr. Substrate preparation as well as the growth process were monitored by *in-situ* reflection high-energy electron diffraction (RHEED) and indicates atomically flat surface. The temperature of the substrate was kept at 280±5 °C for crystalline $Bi_2Se_3$ growth, i.e. in the epitaxial growth regime with a hexagon-on-hexagon orientation and the RHEED pattern of the grown layer is shown in the inset as well. In order to obtain detailed information about the crystal quality of the films, were studied. XRD pattern of the films with for 20 QL $Bi_2Se_3$ show highly resolvable Laue oscillations around the Bragg peaks are a clear indication of the film's structural coherence along the growth direction.

*Band structure calculations*

The ab initio results are obtained with a full-relativistic Korringa-Kohn-Rostoker Green's function method in atomic-sphere approximation [34]. We use the LDA functional by Vosko, Wilk, and Nusair [35]. To identify the character of the states we calculate a k-resolved density (Block spectral density) along the symmetry lines which is additionally resolved by the total angular momentum quantum number j.

**References**

[1] Bovensiepen, U. Femtomagnetism: Magnetism in step with light. *Nat. Phys.* **5**, 461-463 (2009).
[2] Bigot, J.-Y., Vomir, M. & Beaurepaire, E. Coherent ultrafast magnetism induced by femtosecond laser pulses. *Nature Phys.* **5**, 515–520 (2009).
[3] Zhang, G. P., Hübner, W., Lefkidis, G., Bai, Y. & George, T. F. Paradigm of the time-resolved magneto-optical Kerr effect for femtosecond magnetism. *Nature Phys.* 5, 499–502 (2009),
[4] M. Pohl, V.V. *et al.* Ultrafast photoinduced linear and circular optical anisotropy in multiferroic hexagonal manganite $YMnO_3$. *Phys. Rev. B* **88**, 195112 (2013).
[5] Kimel, A. V. *et al.* Room-temperature ultrafast carrier and spin dynamics in GaAs probed by the photoinduced magneto-optical Kerr effect. *Phys. Rev. B* **63**, 235201 (2001).

[6] Ganichev, S.D. *et al.* Spin galvanic effect. *Nature* **417**, 153-156 (2002).

[7] Walser, M.P., Reichl, C., Wegscheider W. & Salis, G. Direct mapping of the formation of a persistent spin helix. *Nature Phys.* **8**, 757–762 (2012).

[8] König, M. *et al.*, Quantum Spin hall insulator in HgTe quantum wells. *Science* **318**, 766-770 (2007).

[9] Fu, L. Kane, C. L. & Mele, E. J. Topological insulators in three dimensions. *Phys. Rev. Lett.* **98**, 106803 (2007)

[10] Zhang H., *et al.* Topological insulators in $Bi_2Se_3$, $Bi_2Te_3$ and $Sb_2Te_3$ with a single Dirac cone on the surface. *Nature Phys.* **5**, 438-442 (2009).

[11] Hosur, P. Circular photogalvanic effect on topological insulator surfaces: Berry-curvature-dependent response. *Phys. Rev. B* **83**, 035309 (2011).

[12] Yokoyama, T. & Murakami S., Spintronics and spincaloritronics in topological insulators. *Physica E: Low-dimensional Systems and Nanostructures* **55**, 1 (2014).

[13] McIver, J. W., Hsieh, D., Steinberg, H., Jarillo-Herrero, P. & Gedik, N. Control over topological insulator photocurrents with light polarization. *Nature Nanotech.* **7**, 96-100 (2011).

[14] Hsieh, D. *et al.* Selective probing of photoinduced charge and spin dynamics in the bulk and surface of a topological insulator. *Phys. Rev. Lett.* **107**, 077401 (2011).

[15] Kastl, C., Karnetzky, C., Karl K. & Holleitner, A. W. Ultrafast helicity control of surface currents in topological insulators with near-unity fidelity. *Nature Comm.* **6**, 6617 (2015).

[16] McIver, J. W. *et al.* Theoretical and experimental study of second harmonic generation from the surface of the topological insulator $Bi_2Se_3$. *Phys. Rev. B* **86**, 035327 (2012).

[17] Junck, A., Refael, G. & von Oppen, F. Current amplification and relaxation in Dirac systems, *Phys. Rev. B* **90**, 245110 (2014).

[18] Sim, S. *et al.* Ultrafast terahertz dynamics of hot Dirac-electron surface scattering in the topological insulator $Bi_2Se_3$, *Phys. Rev. B* **89**, 165137 (2014).

[19] Olbrich P. *et al*. Giant photocurrents in a Dirac fermion system at cyclotron resonance. *Phys. Rev. B* **87**, 235439 (2013).

[20] Olbrich, P. *et al.* Room temperature high frequency transport of Dirac fermions in epitaxially grown $Sb_2Te_3$ and $Bi_2Te_3$ based topological insulators. *Phys. Rev. Lett.* **113**, 096601 (2014).

[21] Wang, Y.H., Steinberg, H., Jarillo-Herrero, P. & Gedik, N. Observation of Floquet-Bloch states on the Surface of a Topological Insulator. *Science* **342**, 453 (2013).

[22] Walowski, J. *et al.* Intrinsic and non-local Gilbert damping in polycrystalline nickel studied by Ti:sapphire laser fs spectroscopy. *J. Phys. D: Appl. Phys.* **41**, 164016 (2008).

[23] Beaurepaire, E., Merle, J.-C., Daunois, A. & Bigot, J.-Y., Ultrafast Spin Dynamics in Ferromagnetic Nickel. *Phys. Rev. Lett.* **76**, 4250 (1996).

[24] Walowski, J., *et al.* Energy equilibration processes of electrons, magnons, and phonons at the femtosecond time scale. *Phys. Rev. Lett.* **100**, 246803 (2008).

[25] Norimatsu, K. *et al.* Coherent optical phonons in a $Bi_2Se_3$ single crystal measured via transient anisotropic reflectivity. *Solid State Comm.* **157**, 58-61 (2013).

[26] Wang, B.T. & Zhang, P. Phonon spectrum and bonding properties of $Bi_2Se_3$: Role of strong spin-orbit interaction. *Appl. Phys. Lett.* **100**, 082109 (2012).

[27] Djordjevic M. *et al.* Comprehensive view on ultrafast dynamics of ferromagnetic films. *Phys. Stat. Sol. C* **3**, 1347 (2006).

[28] Višňovský, Š., Magneto-optical polar Kerr effect and birefringence in magnetic crystals of orthorhombic symmetry. *Czech J. Phys. B* **34**, 155-162 (1984).

[29] Carpene, E., *et al.* Measurement of the magneto-optical response of Fe and $CrO_2$ epitaxial films by pump-probe spectroscopy: Evidence for spin-charge separation. *Phys. Rev. B* **87**, 174437 (2013).

[30] Cote, D., Laman, N. & van Driel, H.M. Rectification and shift currents in GaAs. *Appl. Phys. Lett.* **80**, 905-907 (2002).

[31] Schleicher, J. M., Harrel, S. M. & Schmuttenmaer, C. A. Effect of spin-polarized electrons on terahertz emission from photoexcited GaAs, *J. Appl. Phys.* **105**, 113116 (2009).

[32] Kampfrath, T., *et al.* Terahertz spin current pulses controlled by magnetic heterostructures. *Nature Nanotech.* **8**, 256-260 (2013).

[33] Caho, C. *et al.*, Momentum resolved spin dynamics of bulk and surface excited states in the topological insulator $Bi_2Se_3$. *Phys. Rev. Lett.* **114**, 097401 (2015).

[34] Zabloudil, J., Hammerling, R., Szunyogh, L. & Weinberger, P. Electron Scattering in Solid Matter: A Theoretical and Computational Treatise, Springer Series in Solid-State Sciences, Vol. 147 (Springer, Berlin, 2005).

[35] Vosko, S. H., Wilk, L. & Nusair M. Accurate spin-dependent electron liquid correlation energies for local spin density calculations: a critical analysis. *Can. J. Phys.* **58**, 1200-1211 (1980).

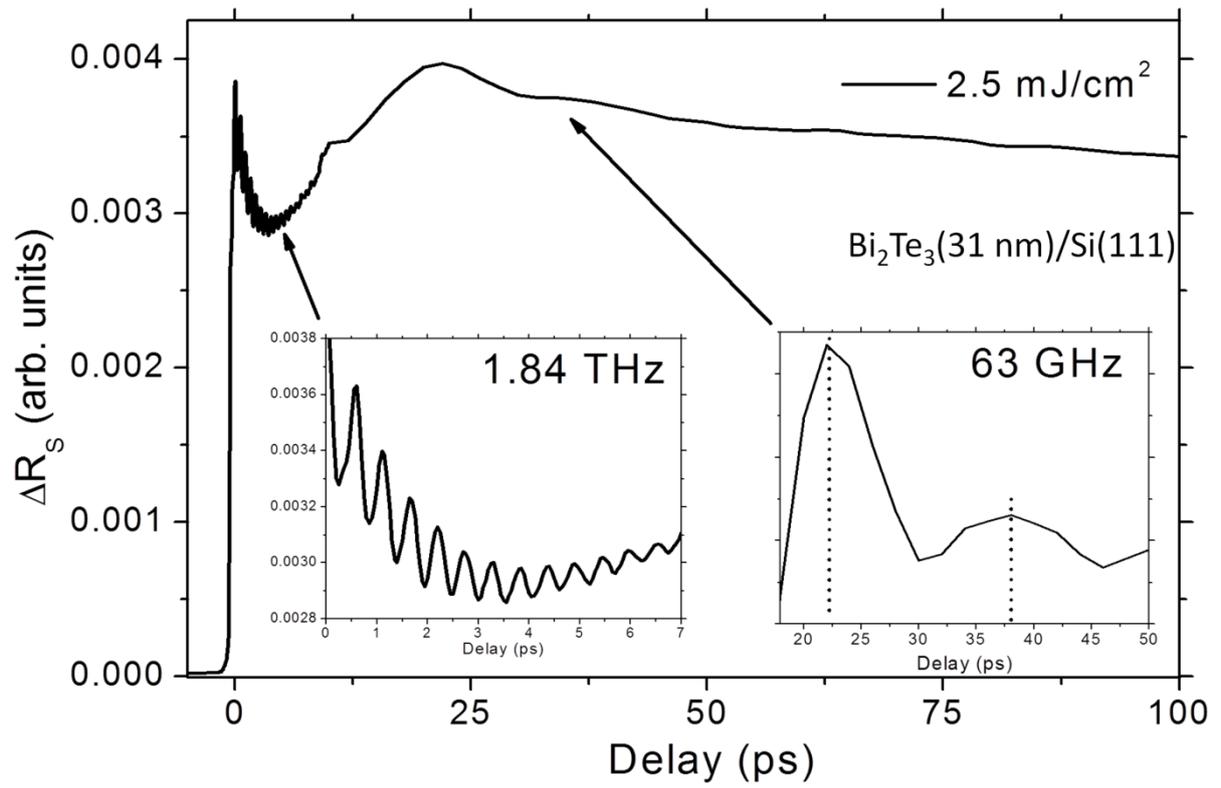

**Fig. 1 Characterization of phonons and stress waves.** Optical phonon mode and acoustic standing wave measured by pump-probe reflectivity dynamics for a $Bi_2Te_3$ film on Si(111) revealing phonon dynamics.

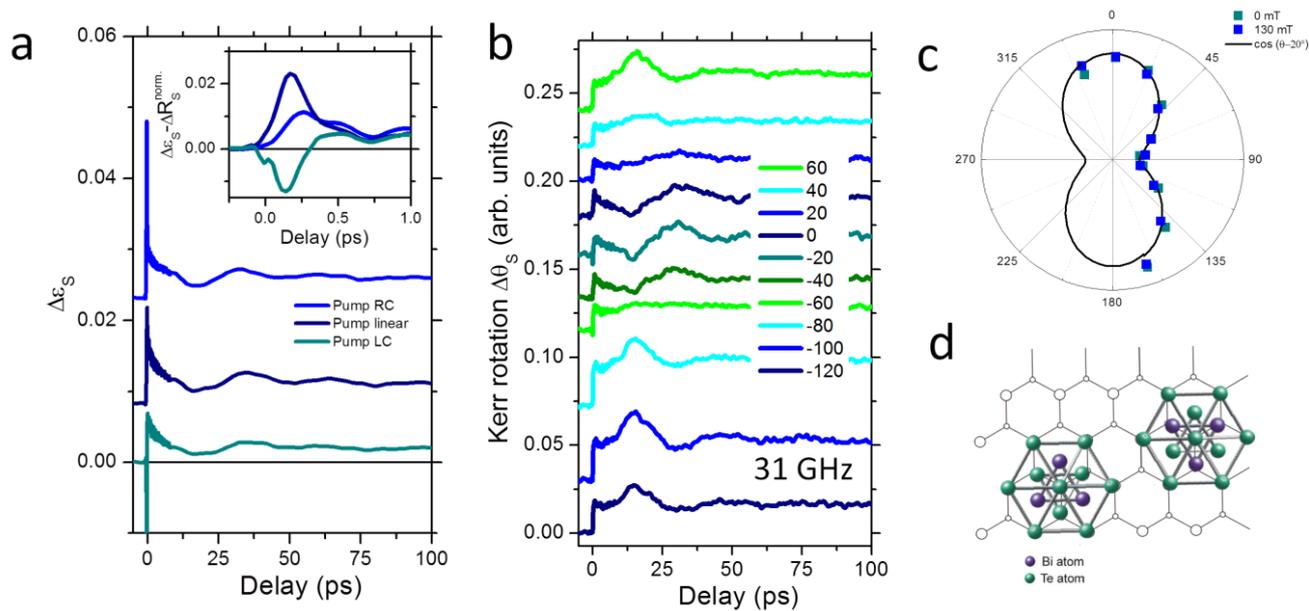

**Fig. 2 Time resolved birefringence.** a) Dynamics on short and long time scales for left circularly (LC), linear and right circularly (RC) polarized pump pulses. Inset: Separation of the coherent signal in the presence of the pump-pulse. b), c) Dynamic mode with 31 GHz with angular dependence of its amplitude and relation to the two domains induced by the Si(111) surface d).

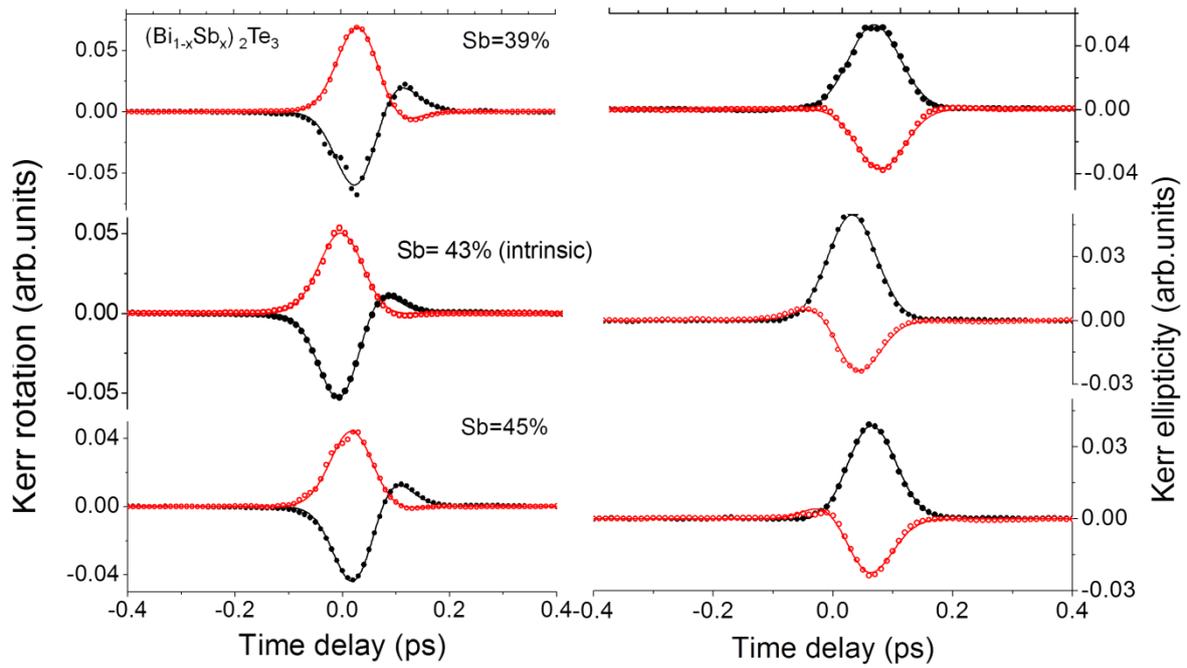

**Fig. 3 Coherent spin signal in presence of the pump-pulse.** Dynamics on short time scales for left circularly (LC, black), and right circularly (RC, red) polarized pump pulses with reference to the linear pump induced changes. The signals are shown for the different compositions of the $(Bi_{1-x}Sb_x)_2Te_3$ film with Sb from p to n doping. Data is overlaid with the model describing dynamic complex Kerr rotation $\theta_K$. Please note that the Kerr ellipticity has a significant delay as compared to the Kerr rotation as expected from the Raman model.

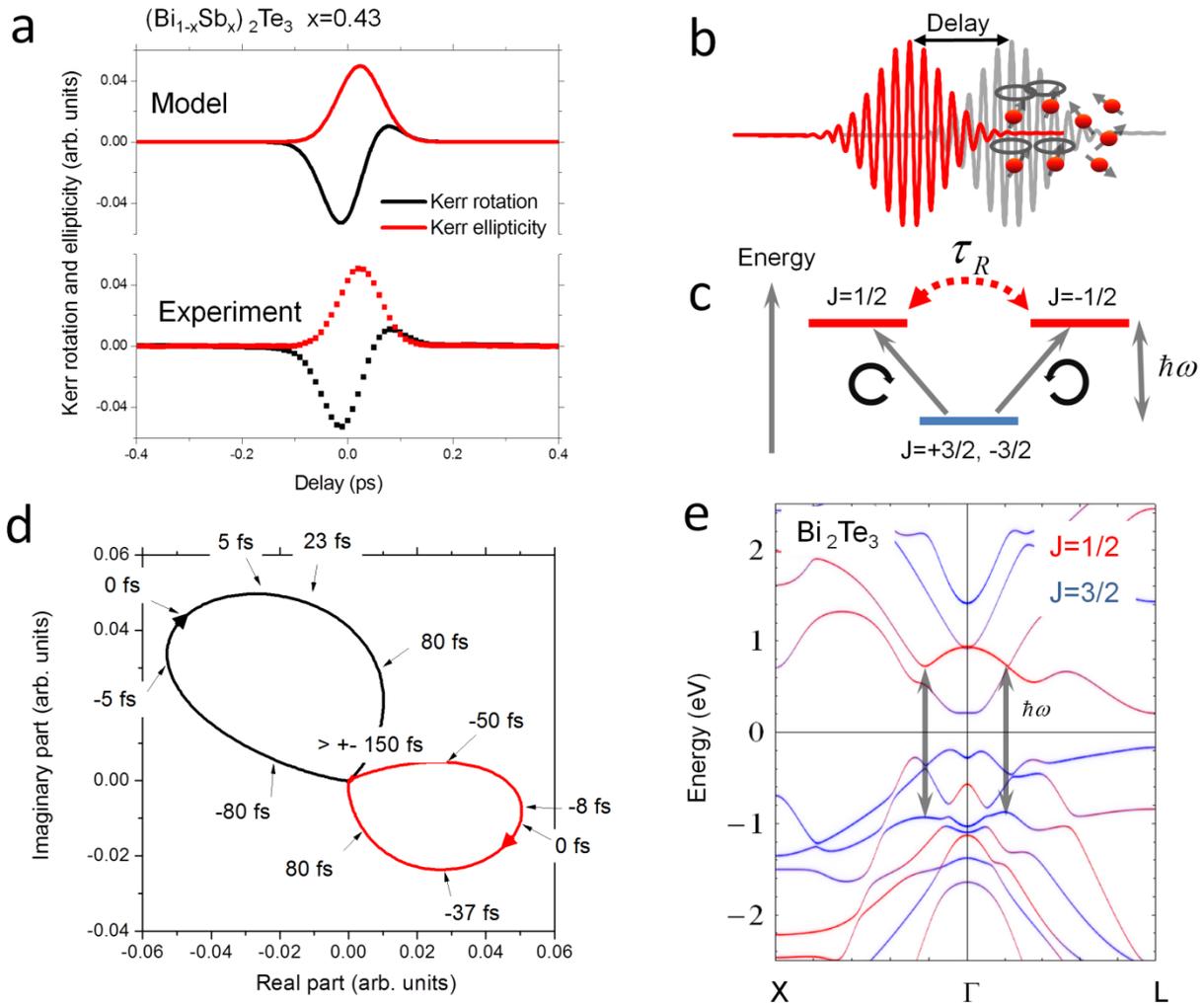

**Fig. 4 Model of the pump-induced coherent polarization at the band edge.** a) Comparison of the dephasing of Kerr rotation and ellipticity, experimental data and model for the compensated case $(Bi_{1-x}Sb_x)_2Te_3$ film with Sb=43%. b), c) Schematics of the coherent interaction in presence of the pump-pulse and in the simplified three level diagram revealing the intrinsic time scale (Raman coherence time $\tau_R$). d) Kerr rotation $\theta_K$ for the compensated case in the complex plane for left circularly (LC, black), and right circularly (RC, red) polarized pump pulses. e) Calculated band structure showing j=1/2, 3/2 character of the states and possible optical transitions for 800 nm (1.55 eV).

# Supplementary Information

**Coherent ultrafast spin-dynamics probed in three dimensional topological insulators**

[1,2]F. Boschini, [1]M. Mansurova, [3]G. Mussler, [3]J. Kampmeier, [3]D. Grützmacher, [4]L. Braun, [5]F. Katmis, [5]J. S. Moodera, [6]C. Dallera, [6]E. Carpene, [7]C. Franz, [7]M. Czerner, [7]C. Heiliger, [4]T. Kampfrath, [8]M. Münzenberg

*1) I. Physikalisches Institut, Georg-August-Universität Göttingen, Friedrich-Hund-Platz 1, 37077 Göttingen, Germany, 2) Dipartimento di Fisica, Politecnico di Milano, 20133 Milan, Italy, 3) Peter Grünberg Institut (PGI-9) and Jülich-Aachen Research Alliance (JARA-FIT), Forschungszentrum Jülich, 52425 Jülich, Germany, 4) Department of Physical Chemistry, Fritz Haber Institute, Faradayweg 4-6, 14195 Berlin, Germany, 5) Massachusetts Institute of Technology, Cambridge, Massachusetts 02139, USA, 6) IFN-CNR, Dipartimento di Fisica, Politecnico di Milano, 20133 Milan, Italy, 7) I. Physikalisches Institut, Justus-Liebig-Universität Gießen, 35392 Gießen, Germany, 8) Institut für Physik, Ernst-Moritz-Arndt Universität Greifswald, Felix-Hausdorff-Straße 6, 17489 Greifswald, Germany*

## 1. Element resolved band structure

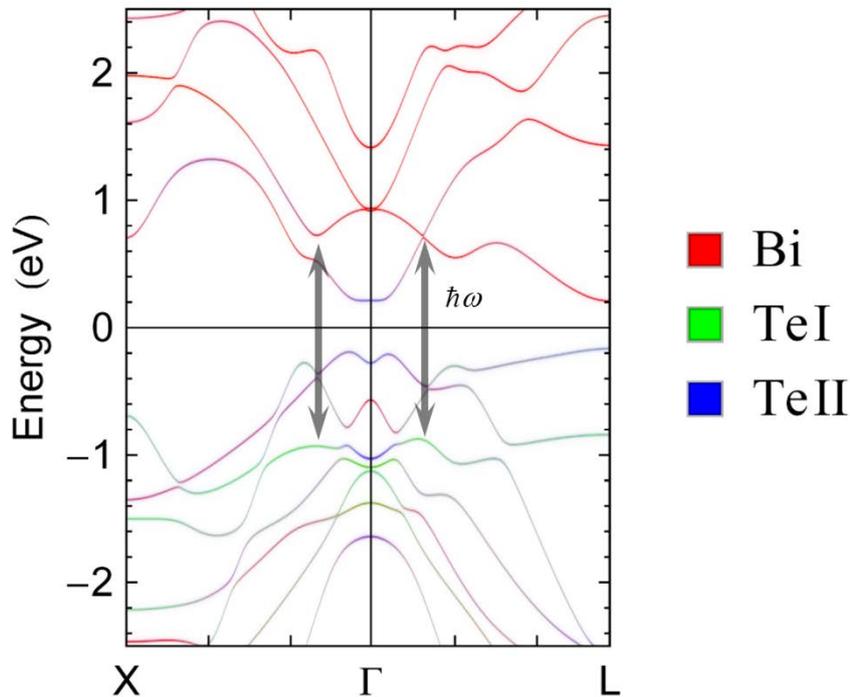

**Fig. 1 Calculated band structure:** showing Bi and Te character of the states that can be related to the optical transitions for the wavelength λ= 800 nm (photon energy 1.55 eV).

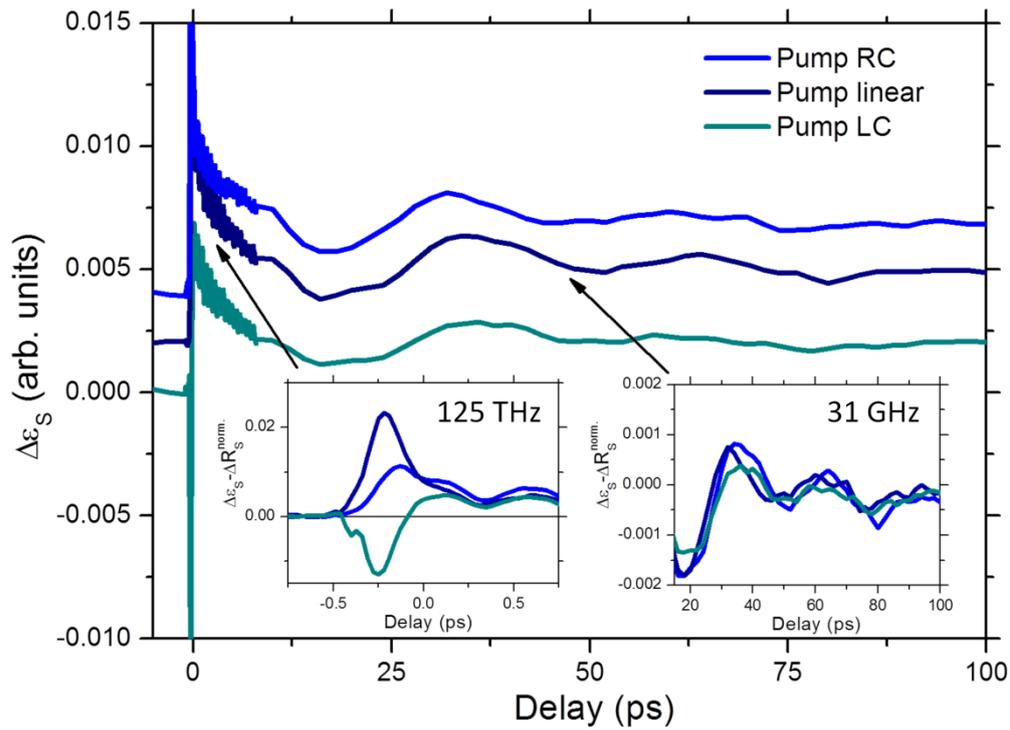

**Fig. 2 Time resolved birefringence** a) Dynamics on short and long time scales for left circularly (LC), linearly and right circularly (RC) polarized pump pulses. Inset left: polarization dependence on the ultrafast time scale. Inset right: the dynamic mode at 31 GHz shows no dependence on the pump-pulse's handedness.